# On the Complexity of Real Functions

Mark Braverman[1]

October 25, 2018

**Abstract**

We develop a notion of computability and complexity of functions over the reals, which seems to be very natural when one tries to determine just how "difficult" a certain function is. This notion can be viewed as an extension of both BSS computability [BCSS98] and bit-computability in the tradition of computable analysis [Wei00] as it relies on the latter but allows some discontinuities and multiple values.

## 1 Introduction

The main goal of this paper is to provide a simple definition which would capture the intuitive notion of whether $f : \mathbb{R} \to \mathbb{R}$ is an "easily" computable function.

There are two main currently existing approaches to the computability and complexity of real functions. One is the BSS approach, where algebraic operations are performed on real numbers that are stored with infinite precision. This approach is described in [BSS89] and [BCSS98]. The other approach, which we call bit-computability, goes all the way back to the Polish school in the 1930-50s, and can be formulated best as: "given a good rational approximation of $x$, compute a good rational approximation of $f(x)$". More recent references on the subject include [Ko91] and [Wei00].

The quality of a given definition for computability of real functions (or any notion of computability in general) can be judged by how well it matches the intuitive notion of "easy" vs. "hard" vs. "impossible". For example, in the discrete case, it it much easier to add two numbers $x + y$ (an operation that takes linear time in the size of the input), than factor an integer $n$ (an operation most believe requires time superpolynomial in the size of the input), and solving the Halting Problem is *truly* impossible (for example, the Goldbach conjecture can be presented as a simple instance of the Halting Problem). We keep this in mind while examining the different approaches to the computability of real functions.

We consider some reservations to both approaches mentioned above. One complaint about the BSS approach is that it is too focused on the algebraic simplicity of the function $f$, allowing only piecewise-semialgebraic functions to be computable. In particular, simple functions such as $x \mapsto e^x$ and $x \mapsto \sqrt{x}$ are not computable in this model (cf. [Brt03b]). There are also problems in the opposite direction: this model classifies all the constant functions $x \mapsto a$, which are extremely simple algebraically, as computable. However, there are infinitely many real $a$'s such that computing $a$ with an arbitrarily good precision would allow us to solve the Halting Problem. We will discuss more of these issues, and ways to 'fix' them in section 3 below.

---
[1]Research is partially supported by an NSERC postgraduate scholarship

On the other hand, in the bit computability model all the computable functions must be continuous. Thus even the simplest step function $s_0(x) = 1$ if $x \geq 0$ and $f(x) = 0$ if $x < 0$ is not computable in this model. This function is extremely simple from the mathematical point of view (involving only the constants 0 and 1 and the comparison operations in its description), hence it would be reasonable to expect it to be computable at least in some sense. For a real function $f$ it is hard to say whether it is uncomputable in this model because it is "difficult" in some profound way, or because it is just discontinuous in one or a few points. It would also be nice to have a setting in which, by analogy with the discrete case, we have a strong connection between the computability of a set $A$ and the computability of its characteristic function $\chi_A$. Another problem is how to properly define the computability of multi-valued functions such as $\sqrt{\ } : \mathbb{C} \to \mathbb{C}$. It is obvious that such a simple function should be computable, but it doesn't have a continuous (and hence bit-computable) branch defined on the entire complex plane $\mathbb{C}$.

We deliberately restrict our discussion to the simple case where the functions are from $\mathbb{R}^k$ or from a simple rectangle such as $[0,1]^k$ to $\mathbb{R}^\ell$. The goal being not to give the broadest definitions and prove the most general theorems, but rather provide a simpler discussion accessible to a broader audience. We will try to use only basic background in the computability and complexity theory and in the topology of $\mathbb{R}^k$.

The paper is organized as follows. In section 2 we give an introduction to the bit model of computation for sets, along with some properties and examples. This section can be viewed as a separate simple introduction to the subject. In section 3 we discuss the BSS computability for sets, and propose three natural modifications to the model that make BSS-computability equivalent to bit-computability. In section 4 we introduce another notion of set-computability – weak computability, and show its equivalence to the standard bit-computability. Sections 3 and 4 can be read independently. In section 5 we use results from sections 3 and 4 to propose a new computability and complexity definition for real functions extending the computability in both models, and the classical complexity in the case of continuous functions.

**Acknowledgement.** The author wishes to thank his graduate supervisor, prof. Stephen Cook, for his insights and encouregement during the preparation of this paper, and for the countless helpful discussions.

# 2 The Bit Model

## 2.1 The Model of Computation

The computability of functions in the bit model as we know it today was first proposed by Grzegorczyk [Grz55] and Lacombe [Lac55]. It has been since developed and generalized. More recent references on the subject include [Ko91], [PR89], and [Wei00].

The basic model of computation here is a Turing Machine. One can think about it as a program in any programming language on the computer. We will usually denote Turing Machines by the letter $M$. Obviously, a computer has a finite memory and cannot store a whole real number. Instead, a *naming system* is used to represent a real number $x$. The most popular naming system for $\mathbb{R}$ uses the *dyadics* $\mathbb{D} = \{\frac{m}{2^n} \mid m \in \mathbb{Z}, n \in \mathbb{N}\}$. A *name* for a number $x$ would be a sequence of dyadics $\phi(1), \phi(2), \phi(3), \ldots$ such that $|\phi(n) - x| < 2^{-n}$. In particular, $\phi(n) \to x$ as $n \to \infty$. Note that the same real number $x$ can have more than one name, e.g. $0, \frac{1}{4}, \frac{2}{8}, \frac{5}{16}, \frac{10}{32} \ldots$ and $\frac{1}{2}, \frac{2}{4}, \frac{3}{8}, \frac{6}{16}, \frac{11}{32} \ldots$ are both names for the same number $x = \frac{1}{3}$.

The oracle terminology is just a natural way to separate the complexity of computing $x$ from the complexity of computing *on* $x$ as a parameter. For most purposes on can think of the oracle $\phi$ for $x$ as an infinite tape containing the binary expansion of $x$.

Consider a function $f : \mathbb{R} \to \mathbb{R}$. In plain language a program $M$ computing $f$ would output $f(x)$, provided the input $x$. More precisely, given a *name* $\phi$ for $x$, $M$ should output a *name* $\psi$ for $f(x)$. Of course, both the name of $x$ and of $f(x)$ take infinitely long to write down in general. We deal with this problem as follows: $M$ is allowed to query $\phi(m)$, which is a $2^{-m}$-approximation of $x$ for any natural $m$, and is required to output $\psi(n)$, where $n$ is given to $M$ as a parameter. In other words, $M$ is allowed to get arbitrarily good approximations of $x$ to compute $f(x)$ with a given precision $2^{-n}$. Note that $M$ does not care about the way $\phi(m)$ is obtained or computed, thus we say that $\phi$ is an *oracle representing* $x$ to $M$, and we write $M^\phi$ to emphasize this fact. We sometimes write $M^\phi(n)$ to emphasize the fact that $M^\phi$ gets one precision parameter $n$.

The definition extends naturally to a function $f : \mathbb{R}^k \to \mathbb{R}^\ell$. Here $M = M^{\phi_1, \phi_2, \ldots, \phi_k}(n)$ is allowed to query each of the $k$ parameters with an arbitrarily good precision and is required to output the $\ell$ values of $f$ with precision $2^{-n}$.

**Example:** Compute the function $f : [0,1] \to [0,1]$, $f(x) = x^2$.

**Solution:** On an input $n$, query for $q = \phi(n+1)$. Output the dyadic number $q^2$.

To show that it works, we need to see that $|q^2 - x^2| < 2^{-n}$. We know that $|q - x| = |\phi(n+1) - x| < 2^{-n-1}$, hence

$$|q^2 - x^2| = |q + x| \cdot |q - x| \leq 2 \cdot |q - x| < 2 \cdot 2^{-n-1} = 2^{-n}.$$

The *running time* of the machine $M^\phi(n)$ is the largest running time over all legitimate oracles $\phi$.

## 2.2 Basic Properties and Examples

One of the main properties of computable functions is that they are continuous. To see this assume that $f$ is computable and let $x$ be a point in the domain of $f$. Let $\phi$ be some oracle representing $x$ as described above. The computation $M^\phi(n)$ terminates after finitely many steps with an output $q$ such that $|q - f(x)| < 2^{-n}$. $M$ only queries $\phi$ with some finite precision $2^{-m}$, and the computation would be the same for any oracle $\psi$ which agrees with $\phi$ on the first $m$ values. If we choose $\phi$ and $\psi$ carefully, we see that $|q - f(y)| < 2^{-n}$ for all $y$ such that $|x - y| < 2^{-m-1}$. This shows that $f$ must be continuous.

Moreover, one can show that if the domain of $f$ is a simple compact (closed and bounded) set, such as the interval $[0, 1]$, then the *modulus of continuity* of $f$ is computable. That is, we can compute a function $m(n)$ such that $|f(x) - f(y)| < 2^{-n}$ whenever $|x - y| < 2^{-m(n)}$.

We recall the definition of computable real numbers, introduced by Turing in [Tur36]. Informally, this definition says that a real number $x$ is computable if we can compute arbitrarily good approximations of $x$.

**Definition 1** *A number $x \in \mathbb{R}$ is computable if and only if a representation $\phi$ of $x$ as described above can be computed.*

Most "standard" continuous functions are computable in this model. For instance, the exponential function $f(x) = e^x$ is computable on $\mathbb{R}$ using the Taylor series expansion of $e^x = \sum_{k=0}^{\infty} \frac{x^k}{k!}$. It is not hard to estimate the number of terms and the precision of $x$ we need to consider in order to get a $2^{-n}$-approximation of $e^x$.

Similarly, polynomial, rational and trigonometric functions with computable coefficients are computable (on properly chosen domains). Moreover, if $f, g$ are computable on the same domain $D$ and $c$ is a computable constant, then $c \cdot f$, $f + g$ and $f \cdot g$ are also computable. $\frac{f}{g}$ is computable if $g \neq 0$ on $A$.

Let $a \in D$ be a computable number, and $f(a) = g(a)$, then the piecewise defined function

$$h(x) = \begin{cases} f(x) & x \in D, \ x \leq a \\ g(x) & x \in D, \ x > a \end{cases}$$

is also computable. Note that the condition $f(a) = g(a)$ is essential here, for otherwise the function $h$ would not be continuous (and cannot be computable in this case).

It must be noted that in the continuous case, only truly "pathological" functions are uncomputable. This notion of computability seems to be effective in classifying *continuous* functions.

One of the possible disadvantages of this model is the fact that even the simpliest discontinuous function, the *step function*

$$s_0(x) = \begin{cases} 0 & x < 0 \\ 1 & x \geq 0 \end{cases}$$

is not computable under this definition. This is in contrary to the intuition that $s_0$ must be a "simple" function. One could also argue that some physical systems, e.g. quantum energy levels, are best described using step functions and other simple discontinuous functions. One of the goals of the present work is to develop notions which deal with this problem.

## 2.3 Complexity of Real Numbers and Real Functions

The time complexity $T_a(n)$ of a number $a$ is the time complexity of computing a $2^{-n}$-approximation of $a$. Note that this is the time the *fastest* machine would take to compute $a$. If $T_a(n) < p(n)$ for some polynomial $p(n)$, then we say that $a$ is poly-time computable.

**Example:** Let's consider the time complexity $T_\pi(n)$ of computing the number $\pi$. One way to compute $\pi$ is to use the Taylor series of arctan:

$$\pi = 4 \cdot \arctan(1) = 4 \cdot \left(1 - \frac{1}{3} + \frac{1}{5} - \frac{1}{7} + \ldots\right) = 4 \cdot \sum_{i=0}^{\infty} \frac{(-1)^i}{2i+1}.$$

It is not hard to see that one would need to take exponentially many terms of the series to obtain a $2^{-n}$-approximation of $\pi$. On the other hand, one can use the Bailey, Borwein and Plouffe [BBP97] formula for $\pi$:

$$\pi = \sum_{i=1}^{\infty} \frac{1}{16^i} \left(\frac{4}{8i+1} - \frac{2}{8i+4} - \frac{1}{8i+5} - \frac{1}{8i+6}\right)$$

(or any one of the many other exponentially convergent $\pi$ expansions). One needs only linearly ($\approx \frac{n}{4}$) terms of the BBP series to get a $2^{-n}$-approximation of $\pi$. Hence $\pi$ is, in fact, poly-time computable.

The definition of the complexity of a function $f$ arises naturally from the definition of computability above. We define the *worst case* time it takes to compute $f(x)$ with precision $2^{-n}$ to be the *time complexity* of $f(x)$, and denote it by $T_f(n)$. Note that $T_f$ may depend on the domain $D$ of $f$. In general, the time complexity may increase as the domain expands.

If $T_f(n) < p(n)$ for some polynomial $p(n)$, we say that the function $f(x)$ is poly-time computable on $D$. The standard functions mentioned in the previous section are typically poly-time computable, using the standard numerical-analytic techniques.

Note that the time complexity of the constant function $f(x) = a$ is equal to the time complexity of the number $a$ (we just ignore the argument $x$).

**Example:** The function $f(x) = e^x$ is poly-time computable on the domain $[0, 1]$. To see this, we use the Taylor series expansion

$$f(x) = e^x = \sum_{i=0}^{\infty} \frac{x^i}{i!}.$$

It is not hard to see that on the $[0, 1]$ interval, $n$ (and even fewer) terms of the series suffice to get a $2^{-n}$-approximation of $f(x)$. All we need to get the sum of the first $n$ terms is $O(n)$ additions and multiplications on $O(n)$-bit numbers. Hence even with the naive $O(n^2)$ multiplication algorithm, we obtain $T_f(n) = O(n^3)$.

## 2.4 Computability and Complexity of Real Sets

According to [BW99], probably, the first definitions of effective subsets of $\mathbb{R}^n$ based on the concept of computability have been proposed by Kreisel and Lacombe in 1957 [KL57], [Lac58]. We refer the reader to [BW99] and [Wei00] for a more detailed discussion. By "computing" a set $C$, we mean generating increasingly precise "images" of $C$. At least for now, we restrict our attention to bounded subsets of $\mathbb{R}^n$.

Consider the two-dimensional case, which is closely related to computer graphics. Intuitively, in this case, the set $C$ is computable if we can draw arbitrarily good "zoom-ins" into it. One can view a $2^{-n}$-precise image of $C$ on a screen as a collection of radius-$2^{-n}$ pixels such that the following two conditions are fulfilled:

1. If a pixel contains a point from $C$, then it is colored black. This ensured that the entire set appears on the screen.

2. If a pixel is far (say $2^{-n}$-far) from $C$, then it is colored white. This ensures that the picture is a faithful image of $C$.

We can take the pixels to be balls of radius $2^{-n}$ with a dyadic center $d \in \mathbb{D}^n$. Formally, we say that $C$ is computable, if there is a machine $M(d,n)$ computing a function from the family

$$f(d,n) = \begin{cases} 1 & \text{if } B(d, 2^{-n}) \cap C \neq \emptyset \\ 0 & \text{if } B(d, 2 \cdot 2^{-n}) \cap C = \emptyset \\ 0 \text{ or } 1 & \text{otherwise} \end{cases} \quad (1)$$

On figure 1 we see some sample values of the function $f$. It should be noted that the definition remains the same if we take square pixels instead of the round ones. It is also unchanged if we replace the ratio between the inner and the outer radius to some $\alpha > 1$ instead of 2.

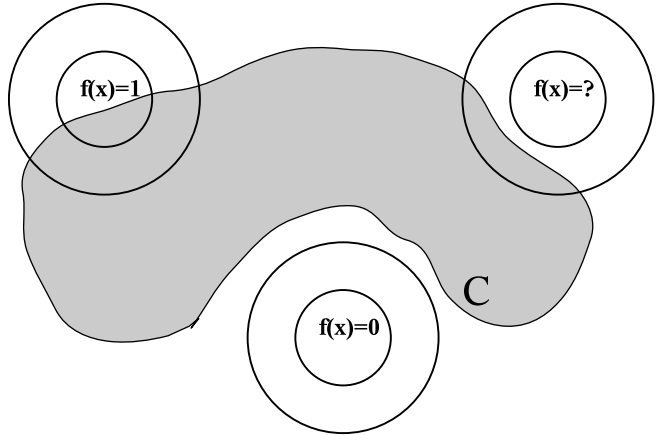

Figure 1: Sample values of $f$. The radius of the inner circle is $2^{-n}$.

One can also define the time complexity $T_C(n)$ as the *worst-case* running time of a machine $M(d,n)$ computing a function from the family (1). Low time complexity means that it is easy to zoom into the set $C$. We say that $C$ is poly-time computable if $T_C(n) < p(n)$ for some polynomial $p$.

So far, one might have been left with the impression that the computability definition above is not very robust. As seen below, on the contrary, it is equivalent to several other reasonable definitions.

First, we establish a connection between the computability of functions and sets. The first attempt is to mimic the discrete case, and to say that a set is computable if and only if its characteristic function $\chi_C$ is computable. Unfortunately, this is not true: a computable function must be continuous, hence $\chi_C$ can only be computable in the trivial cases of $C = \emptyset$ and $C = \mathbb{R}^n$.

The next most natural candidate is the *distance* function

$$d_C(x) = \inf_{y \in C} |x - y|.$$

Unlike the characteristic function, the distance function is continuous and even satisfies the Lipschitz condition $|d_C(x) - d_C(y)| \leq |x - y|$. In fact, we have the following (see [Brv04] for a proof):

**Theorem 2** *1. A bounded set $C$ is computable if and only if the distance function $d_C(x)$ is computable.*

*2. If the distance function is poly-time computable, then the set $C$ is poly-time computable.*

*3. If $n = 1$, the converse to part 2 holds: if the set $C$ is poly-time computable, then $d_C(x)$ is poly-time computable.*

*4. If $n \geq 2$, the converse to part 2 holds if and only if $P = NP$ (which is extremely unlikely).*

Another possible view on the computability of sets it through *global* computability. That is, instead of trying to decide one pixel, we are trying to generate an approximation of the entire set. The word "approximation" here is in the most natural semi-metric on the bounded subsets of $\mathbb{R}^n$, the Hausdorff metric.

The Hausdorff distance between two bounded sets $S$ and $T$ is the smallest quantity $d$ by which we need to "blow-up" $S$ to cover $T$ and vice versa. Formally, for bounded $S$ and $T$, the Hausdorff distance is

$$H(S, T) = \inf\{d \;:\; S \subset B(T, d) \text{ and } T \subset B(S, d)\},$$

where $B(S, d) = \{x \;:\; |x - s| < d \text{ for some } s \in S\}$. On figure 2 is an illustration of the Hausdorff metric.

On the right-hand image the Hausdorff distance is large, because we would need to blow-up $A$ by a large quantity to cover $B$. As a result, the set $B$ is a good picture of $A$ on the left, but not on the right.

The theorem below states that $C$ is computable if and only if one can approximate it in the Hausdorff metric with finite unions of dyadic balls (see [Brv04] for a proof):

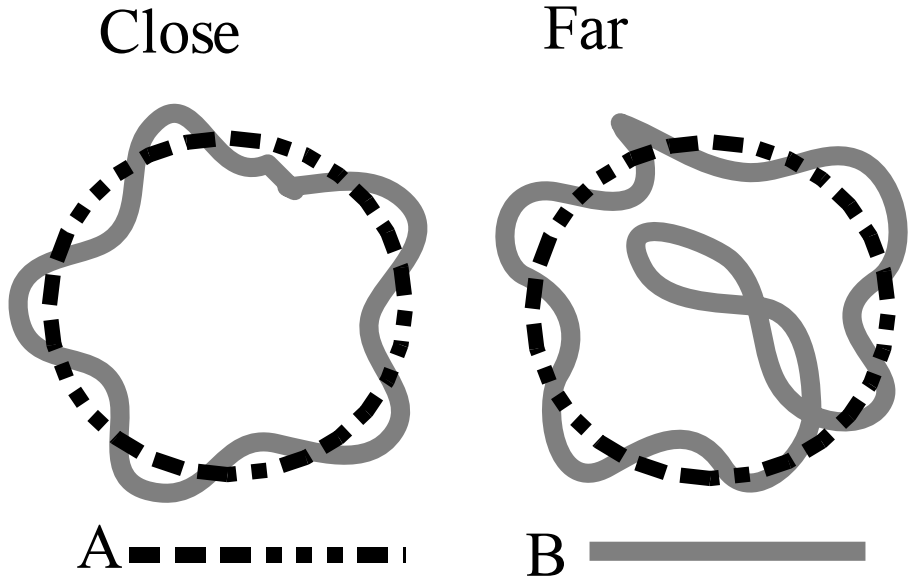

Figure 2: An illustration of the Hausdorff metric

**Theorem 3** *A bounded set $C \subset \mathbb{R}^n$ is computable if and only if there is a machine $M(n)$ that on input $n$ computes a finite sequence of centers $c_1, c_2, \ldots, c_k \in \mathbb{D}^n$ and radii $r_1, r_2, \ldots, r_k \in \mathbb{D}$ such that*
$$H\left(C, \bigcup_{i=1}^{k} \overline{B(c_i, r_i)}\right) < 2^{-n}.$$
*In other words, $\cup_{i=1}^{k} \overline{B(c_i, r_i)}$ is a good Hausdorff approximation of $C$.*

It is not hard to see that when time complexity is concerned, we cannot compare the two definitions. For example $C = [0, 1] \times \{0\} \subset \mathbb{R}^2$ is obviously poly-time computable, but one needs exponentially ($O(2^n)$) many balls to approximate it within $2^{-n}$.

In order to deal with the computability of bounded sets it suffices to discuss compact (= closed and bounded) sets:

**Lemma 4** *A bounded set $C \subset \mathbb{R}^n$ is computable if and only if its closure, $\overline{C}$, is computable.*

We restrict our attention to bounded sets in order for theorem 3 to make sense. We can have the same discussion for unbounded sets by considering the *stereographic projection* of $\mathbb{R}^n$ to the $n$-dimensional sphere.

## 2.5 Examples of Computable Sets

The first natural examples to consider are the simple geometric objects such as a point, a line segment, a circle, etc. For these basic examples set computability connects naturally to number computability:

**Claim 5**  1. A singleton $C = \{c\}$ is computable iff all the coordinates of $c$ are computable numbers.

2. A line segment connecting the points $x$ and $y$ is computable iff all the coordinates of $x$ and $y$ are computable.

3. A ball $\overline{B(c,r)}$ is computable iff $r$ and all the coordinates of $c$ are computable.

This list can be continued with any other standard shapes. In general, such a shape is computable if and only if its parameters are.

There is another connection between computable functions and sets that gives rise to a large family of computable sets. Recall that for a function $f : D \subset \mathbb{R}^k \to \mathbb{R}^\ell$ its *graph* is defined by
$$\Gamma_f = \{(x, f(x)) \mid x \in D\} \subset \mathbb{R}^{k+\ell}.$$
We have the following direct connection between the computability of the function $f$ and of $\Gamma_f$ (as a set).

**Theorem 6** *Suppose $D$ is a computable closed and bounded set, and $f$ is continuous. Then $f$ is computable if and only if $\Gamma_f$ is computable.*

A version of this theorem has been first proven in [Zh96], see also [Brt03a] and [Brv04]. In section 4 we will give a simple proof of it.

In particular, the graphs of all the common continuous functions on closed intervals are computable. E.g. the graph of $f(x) = e^x$ on the $[0, 1]$ interval.

Another interesting family of computable sets are the self-similar fractal images. The most famous set in this family is probably the Cantor set $C$. To define the Cantor set let $C_0 = [0, 1]$ be the $[0, 1]$ interval. Let $C_1$ be the set obtained from $C_0$ by removing its middle third: $C_1 = C_0 \setminus (\frac{1}{3}, \frac{2}{3}) = [0, \frac{1}{3}] \cup [\frac{2}{3}, 1]$. We then remove the middle thirds from each of the two intervals of $C_1$ to obtain $C_2 = [0, \frac{1}{9}] \cup [\frac{2}{9}, \frac{1}{3}] \cup [\frac{2}{3}, \frac{7}{9}] \cup [\frac{8}{9}, 1]$. Continue this process to obtain a chain of closed sets $C_0 \supset C_1 \supset C_2 \supset \ldots$. Define $C = \cap_{i=0}^{\infty} C_i$. See figure 3 for a graphical illustration of the construction.

The Cantor set has a fractal structure because each of its halves is similar to the entire set $C$ with a factor of $\frac{1}{3}$. $C$ has an irrational Hausdorff dimension of $\log_3 2$, which is smaller than 1 but bigger than 0.

We establish that $C$ is computable by showing it is approximable in the Hausdorff metric. $C_i$ is a $3^{-i}$-precise approximation of $C$ in the Hausdorff metric, and it is very easy to approximate $C_i$ in the Hausdorff metric, since $C_i$ is just a union of $2^i$ simple intervals. Thus $C$ is easily computable.

Another famous computable fractal is the *Koch snowflake $K$*. The Koch snowflake is obtained from an equilateral triangle by continuously replacing each side of length $l$ by four sides of length $\frac{l}{3}$, as seen on figure 4. $K$ is a set of Hausdorff dimension $\log_3 4$, which is less than 2 but more than 1. $K$ is the union of three self-similar sets (corresponding to the sides of the original equilateral triangle).

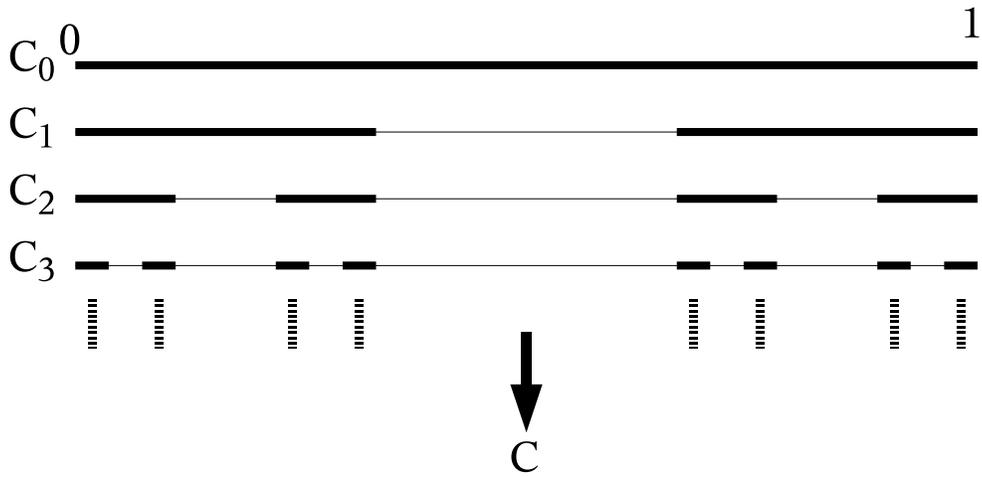

Figure 3: The construction of the Cantor set $C$.

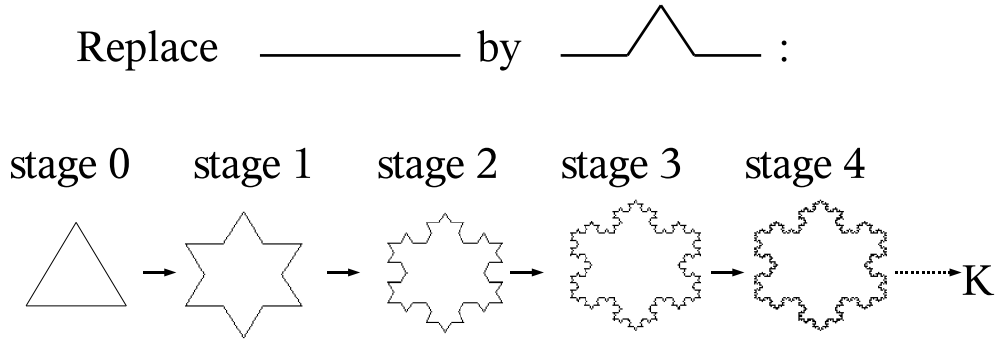

Figure 4: The construction of Koch snowflake $K$.

As in the case of the Cantor set, the $i$-th stage of the construction is a $2^{-\Theta(i)}$-approximation of $K$, and it is easy to compute the $i$-th stage which is a simple union of line segments. This shows that $K$ is Hausdorff approximable, and hence computable.

An intriguing family of computable quasifractals, the hyperbolic Julia sets is discussed in [RW03], [Ret04] and [Brv04]. Other discussions on the computability of Julia sets, and in particular existence of uncomputable Julia sets can be found in [BY04], [BBY04].

# 3 The BSS Model

## 3.1 The Model

The BSS model is quite different from the bit model of computability. Like the bit model, it also extends the standard Turing machines to deal with the continuous reality. In the bit model the extension is through *application* of the standard machine to continuous problems using oracles and naming systems for continuous objects (cf. [Wei00]). The BSS approach extends the *model* of computation itself. We present here an informal description of the model, which is equivalent to a formal one, but is simpler to comprehend for a reader who is new to the subject.

The BSS model in general is formulated for computation over an arbitrary ring or field $R$ (for our purposes one can take $R = \mathbb{R}$ or $R = \mathbb{C}$). The machines in this model are allowed to store an entire element of $R$ in one memory cell. The operations the machine is allowed to perform on numbers are (1) the ring operations ($+$, $-$, $\cdot$, and $\div$ if $R$ is a field); and (2) branching conditioned on exact comparisons between numbers ($=$ and $<$, $\leq$, if $R$ is ordered). Initially, the program is allowed to have only some finite number of constants from $R$. A machine *computes* a function $f : R^k \to R^\ell$ on a domain $D \subset R^k$, if on an input $x \in D$, it outputs $f(x) \in R^\ell$. A machine *decides* a set $C \subset R^n$ if it computes the characteristic function $\chi_C(x) = 1$, if $x \in C$, and $\chi_C(x) = 0$ otherwise.

In the BSS model functions and sets of increasing dimension $n$ are sometimes considered. The underlying structures then can be viewed over $R^\infty = \cup_{n=0}^\infty R^n$. The running time is allowed to depend on $n$. In our discussion we restrict attention to the fixed-dimensional case, but it can be easily extended to $R^\infty$.

In the case $R = \mathbb{Z}_2$ (the finite field with 2 elements) this machine is equivalent to the standard Turing machine. The same statement holds for any finite $R$. In this sense, the BSS model extends the standard computability model. It can be shown that in the case $R = \mathbb{R}$, the model stays the same if we only allow the machine to have finitely many registers (see [BSS89] and [Mich89]).

In general, all the intermediate computation results of a BSS machine $M$, as well as the output, are rational functions $r(x, c)$ of the inputs $x$ and the constants $c$ of $M$. From now on we restrict our attention to $R = \mathbb{R}$. We will need the following definition (see [BCSS98] for more details):

**Definition 7** *A* semi-algebraic *formula $\phi(x_1, \ldots, x_n)$ is a finite combination of polynomial equalities and inequalities over $\mathbb{R}$ linked by the logical connectives $\wedge$ ("and"), $\vee$ ("or"), and $\neg$ ("not").*

Obviously, any semi-algebraic set is computable by a BSS machine (in constant time). As a partial converse, we have the following theorem (cf. Theorem 1 in [BCSS98]).

**Theorem 8** *If a set $C \subset \mathbb{R}^n$ is decided by a BSS machine $M$, then $C$ is a countable disjoint union of semi-algebraic sets.*

A similar statement can be made about BSS computable functions.

## 3.2 Examples of BSS Computable and Uncomputable Sets

The richest family of examples of BSS computable sets are the semi-algebraic sets. In particular, any singleton $C = \{c\}$, any line segment and any ball in $\mathbb{R}^n$ is BSS computable. For example, a ball $\overline{B(x,r)}$ is given by the simple algebraic condition

$$\overline{B(x,r)} = \{y \in \mathbb{R}^n \mid (y_1 - x_1)^2 + (y_2 - x_2)^2 + \ldots + (y_n - x_n)^2 \leq r^2\}.$$

The BSS computable functions include the rational and piecewise rational functions with finitely many computable pieces on BSS computable domains. In particular, the step function $s_0(x)$ which was not computable in the bit model is easily computable in the BSS model: on an input $x$ check whether $x \geq 0$, if yes output 1, otherwise output 0.

In the BSS model, any singleton $\{c\}$ is computable. In particular, if we take $c$ to be some uncomputable number, the singleton $\{c\} \subset \mathbb{R}$ is BSS computable, but not bit-computable. Similarly, the constant function $f(x) = c$ is BSS computable for any $c \in \mathbb{R}$, regardless of whether there is a feasible way to approximate the values of $f$.

We will now present a bit more subtle example of a BSS computable set which is not bit-computable. It will be useful later in the discussion. First, it is well known that there is a computable (binary) predicate $R(x,y)$ such that the predicate $H(x) = \exists y\, R(x,y)$ is uncomputable and if $H(x)$ holds, then the $y$ satisfying $R(x,y)$ is unique. One can think of $x$ as the encoding of a Turing machine, and $R(x,y) = $ "the machine encoded by $x$ halts after exactly $y$ steps". Then $R(x,y)$ is computable by a simple simulation, while $H(x)$ is the halting problem, well known to be undecidable.

| | | R(4,1) | R(3,1) | R(2,1) | R(1,1) |
|---|---|---|---|---|---|
| $I_1$ | ... | | | | |
| $I_2$ | ... | R(4,2) | R(3,2) | R(2,2) | R(1,2) |
| $I_3$ | ... | | | R(2,3) | R(1,3) |
| $I_4$ | ... | | | R(2,4) | R(1,4) |
| | ⋰ | ⋮ | ⋮ | ⋮ | ⋮ |
| | $I_4$ | $I_3$ | $I_2$ | | $I_1$ |

Figure 5: A BSS computable set $C_0$ one cannot draw.

We construct the following closed set $C_0 \subset [0,1] \times [0,1]$. Denote $I_i = [\frac{1}{i+1}, \frac{1}{i}]$ for $i =$

$1, 2, \ldots$. Then $[0,1] = \cup_{i=1}^{\infty} I_i \cup \{0\}$. Define

$$C_0 = (\{0\} \times [0,1]) \cup \bigcup_{R(x,y)=1} I_x \times I_y.$$

It is not hard to see that $C_0$ is closed (there are no accumulation points on the $(0,1] \times \{0\}$ because for each value of $x$ there is at most one value of $y$ such that $R(x,y) = 1$). See figure 5 for a schematic construction of $C_0$.

First we observe that $C_0$ is BSS computable using the following program.

On an input $(x,y) \in [0,1] \times [0,1]$:

1. Check whether $x = 0$; if yes – output 1.

2. Otherwise, check whether $y = 0$; if yes – output 0.

3. Otherwise, find the rectangle(s) $I_i \times I_j$ to which $(x,y)$ belong. This can be done using integer numbers and exact comparisons. There are at most 4 such rectangles, if for one of them $R(i,j) = 1$ – output 1, otherwise – output 0.

Observe that one cannot draw a good image of the set $C$. In fact, in order to decide whether to put a pixel in a small neighborhood of the point $\left(\frac{2i+1}{i(i+1)}, 0\right)$ (this is the middle of the interval $I_i$ on the $x$-axis), one needs essentially to compute $H(i)$, which is impossible.

On the other hand, there are many functions (and sets) that are uncomputable in the BSS model, but are computable in the bit model. For example, the exponential function $f(x) = e^x$ on the interval $[0,1]$ and its graph are uncomputable (see [Brt03b]). In general, one can expect a bit-computable function without some "algebraic structure" to be uncomputable in the BSS model. This is caused by the advantage given to the algebraic operations $(+, -, \cdot$ and $\div)$ in the BSS model.

It is known that BSS computable sets must have an integer Hausdorff dimension (this follows from theorem 8). As a result, the Cantor set (H.d. $\log_3 2$) and the Koch snowflake (H.d. $\log_3 4$) are not computable in the BSS model.

## 3.3 Possible Modifications to the BSS Model

In this section we discuss possible modifications to the BSS models which address some issues raised in the previous section.

Modifications to the model have been discussed in [BV98], [Brt98] leading to a notion of "feasible real RAM" which is essentially equivalent to the bit-computability. The main idea there was, that exact comparisons are not possible on real-life devices, and should not be permitted in the model. This idea, by its nature, bars the step function from being computable. This is not good when one is concerned with classifying functions into "easy" and "hard", or when discussing computability of multi-valued functions.

### 3.3.1 Uncomputable Constants

The first concern to address is the use of uncomputable numbers. It is unreasonable to say that a function $f(x) = a$, where $a$ is a constant encoding the halting problem is computable. The simple solution is to restrict the BSS machines to use only computable constants.

Recall that the computable numbers, which we denote by $\mathcal{C}$, are the numbers that can be approximated arbitrarily well on a computer. It is not hard to see that $\mathcal{C}$ is a countable real closed field, and that $\mathcal{C} + i\mathcal{C} \subset \mathbb{C}$ is algebraically closed. Thus, it makes sense to discuss BSS machines over $\mathcal{C}$ rather than on the entire $\mathbb{R}$. To emphasize the field $\mathcal{C}$ we are working with, we will denote this model by $BSS_\mathcal{C}$.

It now follows easily that simple geometric objects, such as singletons, line segments and balls are $BSS_\mathcal{C}$-computable if and only if they are bit-computable.

### 3.3.2 Computation Errors

The next possible modification addresses the problem of the uncomputability of functions such as $e^x$. The BSS model is in part based on the fact that real-life computers usually use the four arithmetic operations as a base to performing real computations. However, $e^x = \sum_{n=0}^{\infty} \frac{x^n}{n!}$ can be viewed as an infinite-degree polynomial, and is *approximated* arbitrarily well with the finite degree polynomials $p_n(x) = \sum_{i=0}^{n} \frac{x^i}{i!}$. In fact, real-life programs never compute $e^x$, but only $p_n(x)$ with some suitably chosen $n$.

We further modify $BSS_\mathcal{C}$ by allowing the machines to err within a given precision $\varepsilon$. We denote this model by $BSS_\mathcal{C}^\varepsilon$. In this model, a BSS machine $M(x, \varepsilon)$ is said to compute $f(x)$, if on an input $(x, \varepsilon)$, $\varepsilon > 0$, it outputs $f(x)$ with an error of at most $\varepsilon$, and using only computable constants. Note that the simple step function $s_0(x)$ (as well as any other $BSS_\mathcal{C}$ computable functions) is computable in $BSS_\mathcal{C}^\varepsilon$.

One can also define naturally the $BSS_\mathcal{C}^\varepsilon$ computability of sets. A bounded set $C$ is $BSS_\mathcal{C}^\varepsilon$-computable, if there is a BSS machine $M(x, \varepsilon)$ which uses only computable constants, and on input $(x, \varepsilon)$ it outputs 1 if $x \in C$ and 0 if $d(x, C) > \varepsilon$. With this definition, the graph of $e^x$ becomes computable. The simple fractals, such as the Cantor set and Koch snowflake mentioned in section 2.5 also become computable under the modification.

It should be noted that if we drop the requirement of using only computable numbers, all the bounded sets are easily seen to be computable. It is not hard to encode all the "pictures" of any bounded set $C$ into one (possibly uncomputable) number $c$. In other words, all the sets are computable in $BSS^\varepsilon$.

### 3.3.3 Unbounded Computation Branches

The third modification addresses the problem highlighted by the example $C_0$ on figure 5. The problem is the excessive power the BSS model gets from the possibility of having arbitrarily long computation paths (as it happened in the example above).

In the case of "simple" computations, such as $e^x$, or its graph function, we can easily estimate the number of steps the machine would have to perform as a function of $\varepsilon$. We include this condition as an additional restriction on the $BSS_\mathcal{C}^\varepsilon$ machines.

We say that a function (or a set) is $BSS_\mathcal{C}^{\varepsilon,b}$ computable, if it is $BSS_\mathcal{C}^\varepsilon$ computable by a machine $M$, and the running time of $M$ can be bounded by $\tau\left(\lfloor\frac{1}{\varepsilon}\rfloor\right)$ for some integer computable function $\tau : \mathbb{N} \to \mathbb{N}$. We say that $\tau(2^n)$ is the time complexity of the set.

Under this restriction the set $C_0$ on figure 5 is not computable, while the function $e^x$ on $[0,1]$, the graph of $e^x$ and the step function $s_0(x)$ are computable.

As before, if we remove the restriction of using only computable constants, all the sets become computable.

We summarize the modifications to the BSS model in the following diagram:

$$
\begin{array}{ccccc}
BSS_\mathcal{C} & \subset & BSS_\mathcal{C}^\varepsilon & \supset & BSS_\mathcal{C}^{\varepsilon,b} \\
(C_0 \text{ comp., } e^x \text{ not comp.}) & & (C_0 \text{ comp., } e^x \text{ comp.}) & & (C_0 \text{ not comp., } e^x \text{ comp.}) \\
\cap & & \cap & & \cap \\
BSS & \subset & BSS^\varepsilon & \supset & BSS^{\varepsilon,b} \\
(C_0 \text{ comp., } e^x \text{ not comp.}) & & (\text{everything is comp.}) & & (\text{everything is comp.})
\end{array}
$$

We will now show that $BSS_\mathcal{C}^{\varepsilon,b}$-computability is equivalent to bit computability for bounded *sets*. Note that they are still different for *functions*, since the step function $s_0(x)$ is $BSS_\mathcal{C}^{\varepsilon,b}$-computable, but not bit-computable. In section 5 we will connect $BSS_\mathcal{C}^{\varepsilon,b}$ function computability to bit computability.

## 3.4 Computability of sets in $BSS_\mathcal{C}^{\varepsilon,b}$

We show the following:

**Theorem 9** *Let $C \subset \mathbb{R}^k$ be a bounded set. Then $C$ is $BSS_\mathcal{C}^{\varepsilon,b}$-computable if and only if it is bit-computable.*

**Proof:** $C$ is bit-computable $\Rightarrow$ $C$ is $BSS_\mathcal{C}^{\varepsilon,b}$-computable

This is the easier direction. Given a machine $M(d,n)$ for bit-computing $C$, we show how to $BSS_\mathcal{C}^{\varepsilon,b}$-compute it. On an input $\varepsilon$, find $n$ such that $2^{-n} < \frac{\varepsilon}{3}$. Also, by a simple binary search, on an input $x$ to the BSS machine, we can find a dyadic $d \in \mathbb{D}^k$ such that $|x - d| < 2^{-n}$. We can now use the BSS machine to simulate $M(d,n)$. We claim that the output is a valid answer.

If $x \in C$, then $x \in B(d, 2^{-n}) \cap C \neq \emptyset$, and $M(d,n)$ outputs 1. If $d(x, C) > \varepsilon$, then $d(d, C) \geq d(x, C) - |d - x| > \varepsilon - 2^{-n} > \frac{2\varepsilon}{3} > 2 \cdot 2^{-n}$, and $M(d, n)$ outputs 0 in this case.

Finally, we have to bound the running time as a function of $\varepsilon$. This can be done by simulating all the possible runs of $M(d,n)$ with some $n$ such that $2^{-n} < \frac{\varepsilon}{4}$, and all $d$ with denominator of $2^{-(n+k+1)}$.

$C$ is $BSS_\mathcal{C}^{\varepsilon,b}$-computable $\Rightarrow$ $C$ is bit-computable

This is a more involved direction. The reduction we will give is not uniform in $C$. It cannot be uniform due to the fact that $BSS_\mathcal{C}^{\varepsilon,b}$ uses arbitrary computable constants. Even the simple decision questions, such as equality, are not decidable for arbitrary computable reals presented by Turing machines computing them. Denote the $BSS_C^{\varepsilon,b}$ machine computing $C$ by $M(x, \varepsilon)$.

**The nonuniform information needed.** Suppose that the BSS machine $M$ uses $l$ constants $a_1, a_2, \ldots, a_l \in \mathbb{R}$. We would need the following algebraic information about $a_1, \ldots, a_l$, in addition to the Turing machines approximating them:

1. The algebraic degree of $a_i$ over $\mathbb{Q}(a_1, \ldots, a_{i-1})$,

2. if this algebraic degree is finite (i.e. $a_i$ is algebraic over $\mathbb{Q}(a_1, \ldots, a_{i-1})$), the minimal polynomial $p_i(x) \in \mathbb{Q}(a_1, \ldots, a_{i-1})[x]$ with leading coefficient 1, that has $a_i$ as its root. $p_i$ is presented symbolically, with nonleading coefficients given as rational functions with non-zero denominators.

**Lemma 10** *Provided the nonuniform information as above, for any symbolic polynomial $p(x_1, x_2, \ldots, x_l) \in \mathbb{Q}[x_1, x_2, \ldots, x_l]$ we can check whether $p(a_1, a_2, \ldots, a_l) = 0$.*

**Proof:** We prove the lemma by induction on $l$.
**Basis:** For $l = 0$, $p$ is just a rational number, and it is trivial to check whether $p = 0$.
**Step:** Assume it is true for $l = i - 1$, and prove it for $l = i$. Write

$$p(a_1, \ldots, a_{i-1}, a_i) = p_t(a_1, \ldots, a_{i-1})a_i^t + \ldots + p_1(a_1, \ldots, a_{i-1})a_i + p_0(a_1, \ldots, a_{i-1}). \quad (2)$$

There are two cases:
(i) If $a_i$ is algebraic of degree $d$ over $\mathbb{Q}(a_1, \ldots, a_{i-1})$, then we can use the minimal polynomial for $a_i$ to symbolically rewrite (2) as a degree (at most) $d - 1$ polynomial in $a_i$:

$$p(a_1, \ldots, a_{i-1}, a_i) = q_{d-1}(a_1, \ldots, a_{i-1})a_i^{d-1} + \ldots + q_1(a_1, \ldots, a_{i-1})a_i + q_0(a_1, \ldots, a_{i-1}).$$

Here the $q_j$'s are rational functions which have non-zero denominators at $(a_1, a_2, \ldots, a_{i-1})$. By the minimality of $d$, $p(a_1, \ldots, a_{i-1}, a_i) = 0$ if and only if all the $q_j(a_1, \ldots, a_{i-1}) = 0$, $j = 0, 1, \ldots, i - 1$, which we can check by the induction hypothesis (since we know that the denominators are non-zero, all we have to do is check the numerators at $(a_1, a_2, \ldots, a_{i-1})$).

(ii) If $a_i$ is transcendental over $\mathbb{Q}(a_1, \ldots, a_{i-1})$, it follows from (2) that $p(a_1, \ldots, a_{i-1}, a_i) = 0$ if and only if all the $p_j(a_1, \ldots, a_{i-1}) = 0$, $j = 0, 1, \ldots, t$, which we can check by the induction hypothesis. ∎

**Lemma 11** *Provided the nonuniform information as above, for any symbolic polynomial $p(x_1, \ldots, x_l) \in \mathbb{Q}[x_1, \ldots, x_l]$ we can check whether $p(a_1, a_2, \ldots, a_l) > 0$.*

**Proof:** By lemma 10 we can first check whether $p(a_1, a_2, \ldots, a_l) = 0$. If yes, we output 'no'. Otherwise, using increasingly good approximations, we will eventually be able to tell whether $p(a_1, a_2, \ldots, a_l) > 0$ or $p(a_1, a_2, \ldots, a_l) < 0$. ∎

We now return to the proof of theorem 9. Given a dyadic $d \in \mathbb{D}^k$ and $n \in \mathbb{N}$, we would like to compute $f(d, n)$ as in (1):

$$f(d, n) = \begin{cases} 1 & \text{if } B(d, 2^{-n}) \cap C \neq \emptyset \\ 0 & \text{if } B(d, 2 \cdot 2^{-n}) \cap C = \emptyset \\ 0 \text{ or } 1 & \text{otherwise} \end{cases}$$

For the rest of the proof set $\varepsilon = 2^{-n}$. We know there is a computable bound on the running time of $M(x, \varepsilon)$ in terms of $\varepsilon$. We compute this bound $B = B(\varepsilon)$. This means that $M(\bullet, \varepsilon)$ can have at most $2^B$ different computation paths. Each potential path has an output (either 0 or 1), and a set of rational constraints on the input $x = (x_1, \ldots, x_k)$ and the constants $a_1, \ldots, a_l$ that ensure that this path is followed. If the constraints are not satisfiable by any $(x_1, \ldots, x_k)$, it means that the path is never actually followed. The rational constraints can be rewritten as polynomial constraints.

Choose some computation path $\gamma$ on which $M(\bullet, \varepsilon)$ outputs 1. We denote the polynomial constraints to be satisfied in order to follow $\gamma$ by $C_\gamma(x_1, \ldots, x_k, a_1, \ldots, a_l)$. We are interested whether there is an $x \in B(d, 2^{-n}) \cap C$. In particular, we would like to know whether there is such an $x$ that is accepted by the path $\gamma$. This is stated by the following quantified formula

$$f_\gamma(a_1, \ldots, a_l) = \exists x_1, \ldots, x_k \ ((x_1 - d_1)^2 + \ldots + (x_k - d_k)^2 < 2^{-2n}) \wedge C_\gamma(x_1, \ldots, x_k, a_1, \ldots, a_l).$$

Using Tarski's quantifier elimination algorithm (see [Tar51]), we can convert $f_\gamma(a_1, \ldots, a_l)$ into a quantifier-free formula $g_\gamma(a_1, \ldots, a_l)$ which has the same truth value. We then can use lemmas 10 and 11 to decide whether $g_\gamma(a_1, \ldots, a_l)$ is true or false (which is the same as deciding $f_\gamma(a_1, \ldots, a_l)$). A far more efficient procedure can be applied here, using the recent developements in algebraic geometry algorithms and the fact that $f_\gamma$ has only some constant number of existential quantifiers with no alternations. It is possible to reduce the complexity to be exponential in $k$ (and polynomial in the other parameters). See [Ren92] and [BPR03] for the algorithms and their analysis.

As an answer, we output the following

$$f(d, n) = \bigvee_{\gamma \text{ is a 1-valued path of } M(\bullet, \varepsilon)} f_\gamma(a_1, \ldots, a_l).$$

As there are at most $2^B$ such paths $\gamma$, the computation will involve computing $f_\gamma$ at most $2^B$ times.

We now show the correctness of $f(d, n)$. First, suppose that there is an $x \in B(d, \varepsilon) \cap C$ (recall that $\varepsilon = 2^{-n}$). Then $M(x, \varepsilon)$ must output 1. Let $\gamma_x$ be the computation path corresponding to $x$. $\gamma$ is an output-1 path, since $M(x, \varepsilon) = 1$. Thus, we will have $C_\gamma(\vec{x}, a_1, \ldots, a_l) = 1$, and $f_\gamma(a_1, \ldots, a_l) = 1$. So $f(d, n) = 1$ in this case.

Suppose that $f(d, n) = 1$. It means that there is a computation path of $M(x, \varepsilon)$ which accepts an $x \in B(d, \varepsilon)$. Thus, by definition of $BSS_C^{\varepsilon, b}$-computability, there is a $y \in C$ such that $|x - y| < \varepsilon$, and $d(d, C) \leq |d - y| \leq |d - x| + |x - y| < \varepsilon + \varepsilon = 2 \cdot 2^{-n}$. Hence $d(d, C) \geq 2 \cdot 2^{-n}$ implies that $f(d, n) = 0$, which completes the proof. ∎

## 4 Weak Computability of Real Sets

In this section we introduce another computability and complexity notion for subsets of $\mathbb{R}^n$. This notion was first introduced by Chow and Ko in [CK95] under the name of *strong* *recognizability*. It wasn't known at the time that this definition is, in fact, equivalent to the standard bit-computability definition, as we will show below.

The idea of the definition is to relax the conditions of the pixel bit computability definition. We are given a point $x$ as an oracle to $M^\phi(n)$, and we must output 1 if $x \in C$ and 0 if $x$ is $2^{-n}$-far from $C$. Formally,

**Definition 12** *We say that a set $C$ is* weakly *computable if there is an oracle Turing Machine $M^\phi(n)$ such that if $\phi$ represents a real number $x$, then the output of $M^\phi(n)$ is*

$$M^\phi(n) = \begin{cases} 1 & \text{if } x \in C \\ 0 & \text{if } B(x, 2^{-n}) \cap C = \emptyset \\ 0 \text{ or } 1 & \text{otherwise} \end{cases} \tag{3}$$

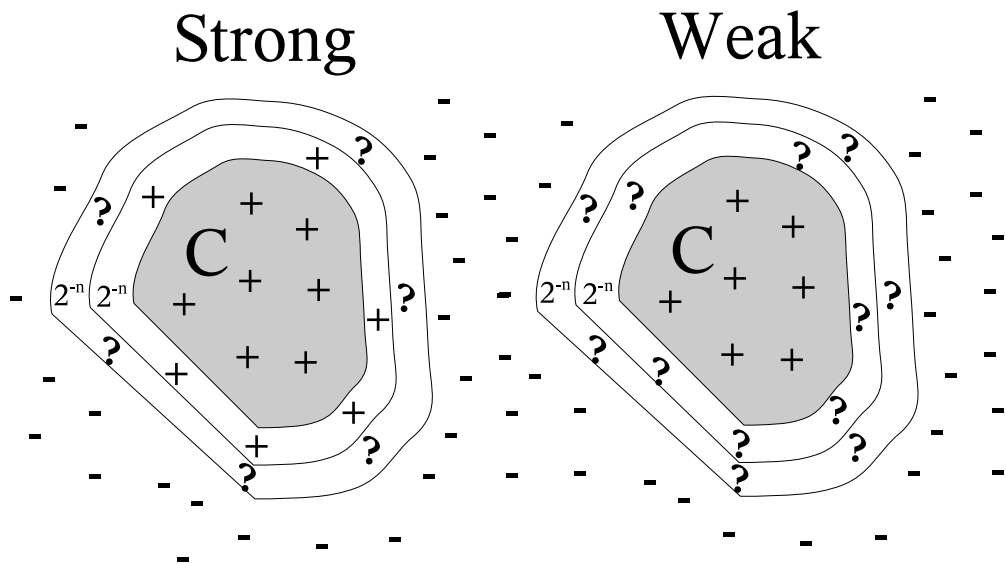

Figure 6: Strong vs. weak computability.

Figure 6 illustrates possible values on input $(x, n)$ for strong (standard) and weak computability. Note that in the case of weak computability $M^\phi(n)$ cannot be computing a function of $x$, because there are no nontrivial continuous $0-1$ valued functions. Hence, for some $x$'es in the 'grey' area the output of $M^\phi(n)$ depends on the specific oracle $\phi$ for $x$.

Our main statement is that, despite the apparent weakness of the weak definition, it is equivalent to the standard bit computability.

**Theorem 13** *A set is bit-computable if and only if it is weakly computable.*

Before proving theorem 13, we will give one application of it to obtain a simple proof of theorem 6 on the equivalence of function and graph computability:

**Theorem 6:** Suppose $D$ is a computable closed and bounded set, and $f$ is continuous. Then $f$ is computable if and only if $\Gamma_f$ is computable.

**Proof:** The "$\Gamma_f$ computable $\Rightarrow f$ is computable" direction is just a routine graph value look-up algorithm and is left to the reader.

We will prove the opposite direction. Namely, that if $f$ is computable $\Rightarrow \Gamma_f$ is *weakly* computable, which implies the computability of $\Gamma_f$ by theorem 13.

Given a point $(x, y)$ by an oracle, we run the $M^\phi$ computing $f$ on $x$ with precision $2^{-(n+2)}$. We respond to all the oracle queries of $M^\phi$ with an answer which is consistent with $x$ as well as some point in $D$. If at any stage of the computation it is impossible, then we've discovered that $x \notin D$, and we output 0. Otherwise, either $x \in D$ and we denote $x' = x$, or there is an $x' \in D$ with $|x - x'| < 2^{-(n+2)}$, such that the run of the machine on $x$ and $x'$ is the same (we don't actually need to know $x'$). In either case, we obtain a value $q$ such that $|q - f(x')| < 2^{-(n+2)}$. If $|q - y| < 2^{-(n+2)}$, output 1; if $|q - y| > 2^{-(n+1)}$, output 0.

To see that the procedure above works, first assume that $(x, y) \in \Gamma_f$. Then $x \in D$ and $y = f(x)$, hence $x' = x$ and $|q - y| = |q - f(x)| < 2^{-(n+2)}$, and we will output 1.

In the opposite direction, if we output 1 then in particular $|x' - x| < 2^{-(n+2)}$ and $|y - f(x')| \leq |y - q| + |q - f(x')| < 2^{-(n+1)} + 2^{-(n+2)} = 3 \cdot 2^{-(n+2)}$. Hence $|(x, y) - (x', f(x'))| \leq |x' - x| + |y - f(x')| < 2^{-n}$, and $d((x, y), \Gamma_f) < 2^{-n}$. Thus we will output 0 whenever $d((x, y), \Gamma_f) > 2^{-n}$. ■

In terms of time complexity, weak computability is provably weaker than the standard computability (assuming $P \neq NP$).

**Theorem 14** *Let $C \subset \mathbb{R}^n$ be a bounded set.*

1. *If $C$ is poly-time bit computable, then $C$ is weakly poly-time computable,*

2. *the converse is equivalent to $P = NP$: "if $C$ is weakly poly-time computable, then it is bit poly-time computable in general" $\Leftrightarrow$ "$P = NP$", which is most unlikely. Hence poly-time bit computability is strictly stronger than weak poly-time computability,*

3. *a weaker version of the converse holds: if $C$ is weakly poly-time computable, then it is exponential time bit computable. Moreover, if the machine which weakly computes $C$ in poly-time reads at most $p(n)$ bits of the input, then $C$ is computable in time $n^{O(1)} 2^{O(p(n)+n)} = 2^{O(p(n)+n)}$.*

See [Brv04] for a proof, which we omit here. The last part of the theorem is useful in proofs of specific upper bounds, e.g. poly-time computability of hyperbolic Julia sets, see [Brv04] for more details.

We will now prove theorem 13.

**Proof:** Recall that bit-computability requires computing a function from the family

$$f(d, n) = \begin{cases} 1 & \text{if } B(d, 2^{-n}) \cap C \neq \emptyset \\ 0 & \text{if } B(d, 2 \cdot 2^{-n}) \cap C = \emptyset \\ 0 \text{ or } 1 & \text{otherwise} \end{cases} \qquad (4)$$

while for weak computability we need to compute a "function" from the family:

$$M^\phi(n) = \begin{cases} 1 & \text{if } x \in C \\ 0 & \text{if } B(x, 2^{-n}) \cap C = \emptyset \\ 0 \text{ or } 1 & \text{otherwise} \end{cases} \quad (5)$$

$C$ is computable $\Rightarrow$ $C$ is weakly computable.

This is the easy direction in the proof. Suppose we have a Turing Machine computing an $f(d, n)$ from the family (4). In order to weakly compute $C$, we first query $d = \phi(n+2)$, and then return $f(d, n+2)$. We need to show that (5) is satisfied.

If $x \in C$, then $|x-d| < 2^{-(n+2)}$ implies that $x \in B(d, 2^{-(n+2)}) \cap C$, so $B(d, 2^{-(n+2)}) \cap C \neq \emptyset$, and by (4) $f(d, n+2)$ returns 1.

If $B(x, 2^{-n}) \cap C = \emptyset$, then $|x - d| < 2^{-(n+2)}$ implies that $B(d, 2^{-n} - 2^{-(n+2)}) \cap C = \emptyset$. So $B(d, 2 \cdot 2^{-(n+2)}) \cap C \subset B(d, 2^{-n} - 2^{-(n+2)}) \cap C = \emptyset$, and by (4) $f(d, n+2)$ returns 0, which completes the proof.

$C$ is weakly computable $\Rightarrow$ $C$ is computable.

We will show the implication for a one-dimensional set $C$, the proof in $k > 1$ dimensions works out in a similar fashion. For convenience purposes assume that $C \subset \left[\frac{1}{4}, \frac{3}{4}\right]$, and hence in (4) we only need to consider $d \in [0, 1]$. The proof extends trivially to bigger intervals ($C$ is bounded).

We construct an infinite tree $T$. In every vertex of $T$ we write a dyadic number. The numbers on level $l$ are dyadics of the form $m \cdot 2^{-l}$. The root, which is on level 1, is labeled by $\frac{1}{2} = 0.1$ (all the numbers in this section are in binary notation). Each vertex $v$ on a level $l$ has 3 children. If the label of $v$ is $m \cdot 2^{-l}$ then the labels of its children are $m \cdot 2^{-l} - 2^{-(l+1)}$, $m \cdot 2^{-l}$ and $m \cdot 2^{-l} + 2^{-(l+1)}$, or in other words $(2m-1) \cdot 2^{-(l+1)}$, $2m \cdot 2^{-(l+1)}$ and $(2m+1) \cdot 2^{-(l+1)}$. On figure 7 we see the first three levels of the tree (cf [Wei00], section 7.2, signed digit representation).

It is easy to see that numbers on every path $p$ in the tree converge to a real number $x_p \in [0, 1]$. Conversely, for every $x \in [0, 1]$ there is a path $p$ such that $x_p = x$ (e.g. choose $p$ to be the prefixes of the binary expansion of $x$). Moreover, if we denote by $p(n)$ the label of the $n$-th node on $p$, then $|x - p(n)| \leq 2^{-n} < 2^{-(n-1)}$. Hence $\phi(n) = p(n+1)$ is a valid oracle for $x_p$.

We will now describe how to compute $f(d, n)$ as in (4). On an input $(d, n)$, find two nodes $v_1$ and $v_2$ on level $n$ such that $\text{label}(v_1) \leq d \leq \text{label}(v_2)$ and $|\text{label}(v_1) - d| < 2^{-n}$, $|\text{label}(v_2) - d| < 2^{-n}$ (if $d$ is an integer multiple of $2^{-n}$, we can choose $v_1 = v_2$). Denote the paths from the root to $v_1$ and $v_2$ by $p_1$ and $p_2$, respectively. Denote the machine weakly computing $C$ by $M^\phi(n)$. We simulate the computation of $M^\phi(n)$ on the subtrees with roots $v_1$ and $v_2$ as follows.

Consider the simulation with root $v_1$ (the simulation with root $v_2$ is done in the same way). For every oracle query $\phi(m)$ with $m < n$, we return the value of $p_1(m+1)$ (which is a valid output for the oracle). Otherwise we consider all the possible descendants of $v_1$ on level $m + 1$, and create a separate computation for each of them (thus creating $3^{m-n+1}$ computations). Consider one of the copies and denote the path leading to the selected vertex on level $m+1$ by $p$ ($p$ extends $p_1$). If we are now asked about $\phi(r)$ for some $r < m+1$, we return the value of $p(r+1)$, and otherwise we again consider all possible descendants of $p(m+1)$ on level $r+1$, and split the computation into $3^{r-m}$ computations. We continue this process until all computations terminate.

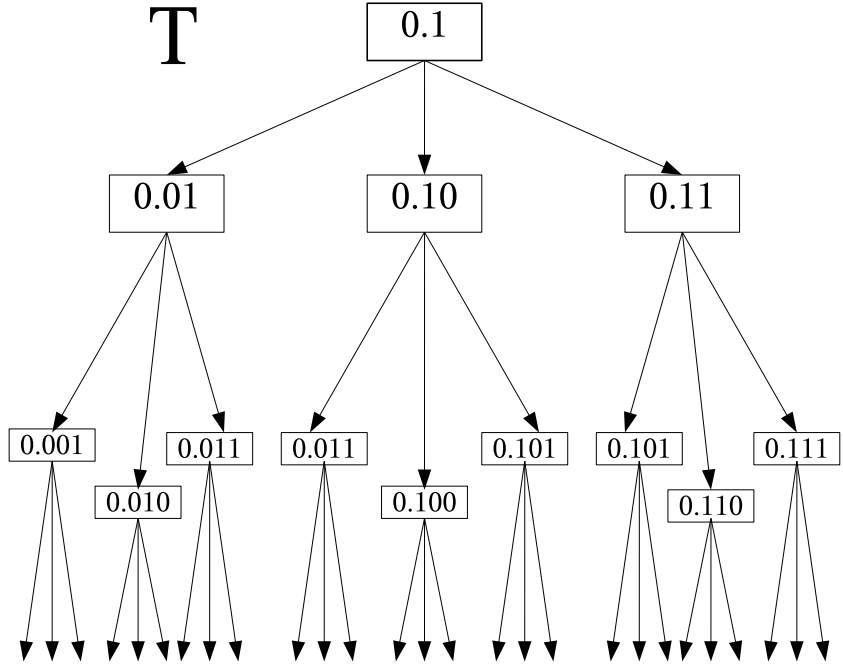

Figure 7: The first three levels of the tree $T$

If any one of the computations (starting either from $v_1$ or from $v_2$) returns 1, we return 1. Otherwise return 0. We need to show two things:

1. The algorithm terminates.

2. It gives answers that satisfy (4).

We will be using König's lemma.

**König's Lemma:** If a every vertex in a tree has a finite degree, then the tree is infinite if and only if it has an infinite branch.

Suppose that the computation does not terminate. We can view the entire computation as a tree where the nodes are the subcomputations described above and a computation $C_1$ is the parent of the $3^s$ computations it launches. If the entire computation does not terminate, then there are two possibilities: either one of the computations $C'$ fails to terminate without calling to subcomputations, or the tree of all the computations to be performed is an infinite tree.

In the first case denote the path in $T$ leading to $C'$ by $p'$. Then $p'$ corresponds to a dyadic number $d'$, and also gives an oracle $\phi'$ for $d'$. Note that $C'$ is reached and executed by simulating $M^{\phi'}(n)$. Hence $M^{\phi'}(n)$ does not terminate in this case, contradiction.

In the second case, by König's lemma, there must be an infinite branch in the computations tree. Denote the branch by $C_1, C_2, C_3, \ldots$. That is, $C_1$ calls $C_2$, $C_2$ calls $C_3$ etc. Note that each $C_i$ works with a path $p_i$ of $T$ and $p_{i+1}$ strictly extends $p_i$ for each $i$, hence the infinite sequence of $C_i$ corresponds to an infinite path $p$ in $T$. The path converges to a real number $x \in [0,1]$, and $p$ gives rise to an oracle $\phi$ for $x$. By the construction, the sequence of $C_1, C_2, C_3, \ldots$ simulates the computation of $M^\phi(n)$. Hence $M^\phi(n)$ does not terminate, contradiction. This shows that the algorithm terminates.

We now have to show the correctness of the algorithm.

**Case 1:** $B(d, 2^{-n}) \cap C \neq \emptyset$. In this case, either $v_1$ or $v_2$ has a descending path $p$ in $T$ which converges to an $x \in C$. Consider the oracle $\phi$ corresponding to this path. One of the computation paths of the algorithm will be a simulation of $M^\phi(n)$, and hence will have to output 1.

**Case 2:** $B(d, 3 \cdot 2^{-n}) \cap C = \emptyset$. In this case all points corresponding to descendants of $v_1$ and $v_2$ are at distance at most $2 \cdot 2^{-n}$ from $d$, and hence are at least $2^{-n}$-far from $C$. Hence any computation corresponding to simulating $M^\phi$ along any of the paths must output 0.

Note that we are only able to compute a function satisfying a condition weaker than (4). Namely, we compute a function $f$ such that

$$f(d,n) = \begin{cases} 1 & \text{if } B(d, 2^{-n}) \cap C \neq \emptyset \\ 0 & \text{if } B(d, 3 \cdot 2^{-n}) \cap C = \emptyset \\ 0 \text{ or } 1 & \text{otherwise} \end{cases}$$

It is very easy to use $f$ to compute a function that satisfies (4). Take

$$g(d,n) = f(d - 2^{-(n+1)}, n+1) \vee f(d, n+1) \vee f(d + 2^{-(n+1)}, n+1)$$

If $|d - c| < 2^{-n}$ for some $c \in C$, then either $|d - c + 2^{-(n+1)}| < 2^{-(n+1)}$, or $|d - c| < 2^{-(n+1)}$, or $|d - c - 2^{-(n+1)}| < 2^{-(n+1)}$. So one of the $f$'s will return 1. On the other hand, if $B(d, 2 \cdot 2^{-n}) \cap C = \emptyset$, then $B(d - 2^{-(n+1)}, 3 \cdot 2^{-(n+1)}) \cap C = \emptyset$, $B(d, 3 \cdot 2^{-(n+1)}) \cap C = \emptyset$ and $B(d + 2^{-(n+1)}, 3 \cdot 2^{-(n+1)}) \cap C = \emptyset$. Hence $g$ returns 0 in this case. This completes the proof. ∎

## 5 Complexity of Real Functions

In this section we propose a new definition for the computability and complexity of real functions, and establish its connections to bit and BSS computability.

### 5.1 Computability of Real Functions

The main idea arises from the equivalence between the function computability and the set computability in case of a continuous function, which was established in theorem 6. We have a good notion of bit-computability for sets which coincides with $BSS_\mathcal{C}^{\varepsilon,b}$-computability. We use it to define a computability notion for functions:

**Definition 15** *We say that a bounded real function $f$ on a bounded domain from a class $\mathcal{F}$ is* graph computable, *if its graph $\Gamma_f$ is computable as a set.*

Obviously, one has to be careful about the choice of the class $\mathcal{F}$ which extends the class of the continuous functions. For example, under this definition, any function $f : [0,1] \to [0,1]$ which has a dense graph in $[0,1] \times [0,1]$ would be computable under this definition. On the other hand, one can reasonably interpret this example by saying that all we can know about $f(x)$ in this case is that it is some number in $[0,1]$ (regardless of the precision $x$ is given with), and that is exactly what we can read from the graph.

The definition above extends naturally to multi-valued functions. One possible candidate for the class $\mathcal{F}$ is the class of (multi-valued) functions $f : D \subset \mathbb{R}^n \to \mathbb{R}^m$ such that $f$ has at most finitely many limits at any point of $D$. In other words,

$$|\{x\} \times \mathbb{R}^m \cap \overline{\Gamma_f}| < \infty \tag{6}$$

for all $x \in D$. We denote this class by $\mathcal{F}_f$. $\mathcal{F}_f$ extends the class of the continuous functions. Note that both the step function and the square root function on $\mathbb{C}$, viewed as a function from $\mathbb{R}^2$ to $\mathbb{R}^2$, belong to this class. In the case of a function $f : \mathbb{R}^1 \to \mathbb{R}^1$, all the $f$'s of bounded variation satisfy condition (6). The definition will make sense outside the class $\mathcal{F}_f$.

By theorem 6, definition 15 coincides with the bit computability definition in the case of continuous functions on closed domains. In the next section we investigate its connection to the BSS model, and show that it extends the $BSS_\mathcal{C}^{\varepsilon,b}$-computable functions.

## 5.2 Connections to the BSS model

In this section we show that all the $BSS_\mathcal{C}^{\varepsilon,b}$-computable functions are graph-computable. The converse is not true. For example, let $\Gamma_e$ be the graph of the exponential function $e^x$ on $[0,1]$. The characteristic function $\chi_{\Gamma_e}(x)$ is easily seen not to be $BSS_\mathcal{C}^{\varepsilon,b}$ computable, even though it is graph computable.

**Theorem 16** *If a function $f : \mathbb{R}^k \to \mathbb{R}^\ell$ is $BSS_\mathcal{C}^{\varepsilon,b}$-computable and bounded on some bounded semi-algebraic $BSS_\mathcal{C}$-computable domain $D$ (e.g. $D = [a_1, b_1] \times [a_2, b_2] \times \ldots \times [a_k, b_k]$ with computable endpoints), then it is graph computable.*

**Proof:** We first note that if a function is $BSS_\mathcal{C}^{\varepsilon,b}$-computable on $D$ by some machine $M(z, \varepsilon)$, then its graph is $BSS_\mathcal{C}^{\varepsilon,b}$-computable. To see this let $(x, y) \in \mathbb{R}^{k+\ell}$ and $\varepsilon > 0$ be given. First of all check whether $x \in D$, and if not – reject. Next, simulate the run of $M(x, \frac{\varepsilon}{2})$ to obtain $z$, $|z - f(x)| < \frac{\varepsilon}{2}$. If $|y - z| < \frac{\varepsilon}{2}$, output 1, otherwise output 0.

If $y = f(x)$, then $|z - y| < \frac{\varepsilon}{2}$ and we will output 1. If $d((x,y), \Gamma_f) > \varepsilon$, then in particular $|y - f(x)| > \varepsilon$, so $|y - z| \geq |y - f(x)| - |f(x) - z| > \varepsilon - \frac{\varepsilon}{2} = \frac{\varepsilon}{2}$, and the output is 0. It is also not hard to see that the running time is bounded by a computable function of $\varepsilon$.

We use theorem 9 to conclude that the graph $\Gamma_f$ is bit-computable, so $f$ is graph-computable. ∎

## 5.3 Complexity of Real Functions

By now we have established that graph computability is a useful notion extending both bit computability and a natural modified version of the BSS computability. Our next goal is to give a reasonable definition for *graph complexity* of real functions. Should we take the complexity of computing $\overline{G_f}$? Or the complexity of weakly computing it? It turns out that neither one extends the bit-complexity of $f$ in the continuous case. In fact, we have the following theorem.

**Theorem 17** *Let $f : [0,1]^k \to [0,1]^\ell$ be a continuous function, with a polynomial modulus of continuity $\mu(n)$ ($|f(x) - f(y)| < 2^{-n}$ whenever $|x - y| < 2^{-\mu(n)}$). Denote the following properties:*

(a) *The graph $\Gamma_f$ is poly-time computable as a set.*
(b) *$f$ is a poly-time computable function.*
(c) *The graph $\Gamma_f$ is weakly poly-time computable as a set.*
*Then we have the following:*

1. $(a) \Rightarrow (c)$ *and* $(b) \Rightarrow (c)$,

2. *in general,* $(b) \Rightarrow (a)$ *implies* $P = NP$, *and*

3. $(a) \Rightarrow (b)$ *(and also* $(c) \Rightarrow (b)$*), implies that integer factoring and other one-way functions can be done in poly time (this is weaker than $P = NP$, yet extremely unlikely).*

**Proof:** The first claim is quite routine and is left to the reader.

We sketch the proof of the second claim. The goal is to construct a poly-time computable function on the $[0, 1]$ interval for which it is NP-hard to strongly compute its graph. The idea is to subdivide $[0, 1]$ into the intervals $I_i = \left[\frac{1}{i+1}, \frac{1}{i}\right]$. Some of them correspond to boolean formulas $\phi_i$. For such an $i$ we subdivide the middle fifth of $I_i$ into $2^k$ subintervals, where $k$ is the number of variables in $\phi_i$. We associate a spike of $f$ to each thruth assignment satisfying $\phi_i$. Then it is easy to compute $f$ in poly-time using only one substitution to $\phi_i$, but a strong query to $\Gamma_f$ would allow us to check for the satisfiability of $\phi_i$ fast.

We will now prove the third claim, namely that $(a) \Rightarrow (b)$ would allow fast factoring of integers. The example we construct is for $k = 1$, $\ell = 2$. It is unclear whether such an example can be constructed for $k = \ell = 1$.

The idea of the proof is as follows. We construct a function $f$ which has the values of some one-way function $F$ encoded into it, so that computing $f$ is as hard as computing $F$. We make the graph of $f$ "easy" to draw by the following construction. Let $C$ be a (small) interval on which the value $F(n)$ is encoded. We make $f$ fairly dense on intervals $CL$ and $CR$ lying on both sides of $C$. So that its graph in coarse precision looks like two "walls" on $CL$ and $CR$, and there is no need to worry about the value on $C$ to draw the graph. In fince precision, on the other hand, the pixels we can query are so small that to answer them on $C$ we would only need to know whether $F(n) = k$ for a few different $k$'s, which is easy to check.

We now give a more formal construction. First consider the following function $F : \mathbb{N} \to \{0,1\}^*$: on an input $n$, $F$ outputs a canonic encoding of the prime factorization of $n$. E.g. $F(6) = <2,3>$, $F(24) = <2,2,2,3>$, all encoded properly as binary strings. Then $F$ satisfies the following properties:

- $F$ is well-defined on all $\mathbb{N}$,
- $F$ is one-to-one,
- the length of $F(n)$ is linear in the length of $n$: $|F(n)| < c \cdot |n|$, and
- given a string $s \in \{0,1\}^*$ and a number $n$ it is easy (poly-time) to verify whether $F(x) = s$.

Subdivide the $[0,1]$ interval into the intervals $I_j = \left[\frac{1}{j+1}, \frac{1}{j}\right]$. We subdivide each $I_j$ into 5 intervals: left – $L_j$, central-left – $CL_j$, central – $C_j$, central-right – $CR_j$ and right – $R_j$. Let $k = \lceil \log j \rceil$. The dividing points of the intervals are dyadic, and the lengths are as follows:

$$Lnth(L) \approx Lnth(R), \quad Lnth(CL) = Lnth(C) = Lnth(CR) = 2^{-(ck+k+3)} \ll Lnth(L).$$

Denote
$$CL = [x_{ll}, x_{lr}], \quad C = [x_{lr}, x_{rl}] \quad CR = [x_{rl}, x_{rr}].$$

$f$ on $CL$ is defined to have the following properties:

(a) the range of $f$ is in $[-2^{-k}, 2^{-k}] \times [-2^{-k}, 2^{-k}]$,
(b) $f(x_{ll}) = f(x_{lr}) = (0,0)$,
(c) for any $(x,y)$ such that $|x|, |y| < \alpha < 2^{-k}$ there is a $z \in CL$ such that $|z - x_{lr}| < \alpha \cdot 2^{-(ck+3)}$ and $|f(z) - (x,y)| < \alpha \cdot 2^{-(ck+k+3)}$,
(d) $f$ is poly-time graph computable on this interval.

This can be achieved, for example, by a function presented on figure 8 (left).

Next, we define $f$ on $C$ as follows:

$$f(z) = \begin{cases} (2^{ck} \cdot (z - x_{lr}), F(n) \cdot (z - x_{lr})), & \text{if } z \text{ is in the left half of } C \\ (2^{ck} \cdot (x_{rl} - z), F(n) \cdot (x_{rl} - z)), & \text{if } z \text{ is in the right half of } C \end{cases}$$

See figure 8 (right). Note that $f(x_{lr}) = f(x_{rl}) = (0,0)$. We define $f$ on $CR$ symmetrically to its definition on $CL$ above. $f$ is defined to be $(0,0)$ on the intervals $L$ and $R$. It is easy to see now that $f$ extends continuously to the closed interval $[0,1]$ by setting $f(0) = (0,0)$.

If $f$ were poly-time computable, then querying its value on $C$ for the interval $I_n$ with precision $2^{-O(\log n)}$ would allow us to compute $F(n)$. Our goal now is to show that the graph $\Gamma_f$ is poly-time computable as a set. The idea is as follows: if we query a fairly large pixel of $\Gamma_f$, then the picture will be dominated by (the computationally easy) graph of $f$ on $CL$ and $CR$, and we will not have to worry about the value of $f$ on $C$. If we query a fairly small pixel in $C$, then to determine the answer we will have to verify at most one potential value of $F$, which can be done in poly-time. On an input $(d,m)$, $d = (z, (x,y)) \in \mathbb{D}^3$, we will decide whether $B(d, 2^{-m}) \cap \Gamma_f \neq \emptyset$ or $B(d, 8 \cdot 2^{-m}) \cap \Gamma_f = \emptyset$, with the other cases being the "grey area". It is not hard to see that this is equivalent to the standard definition of computing a set.

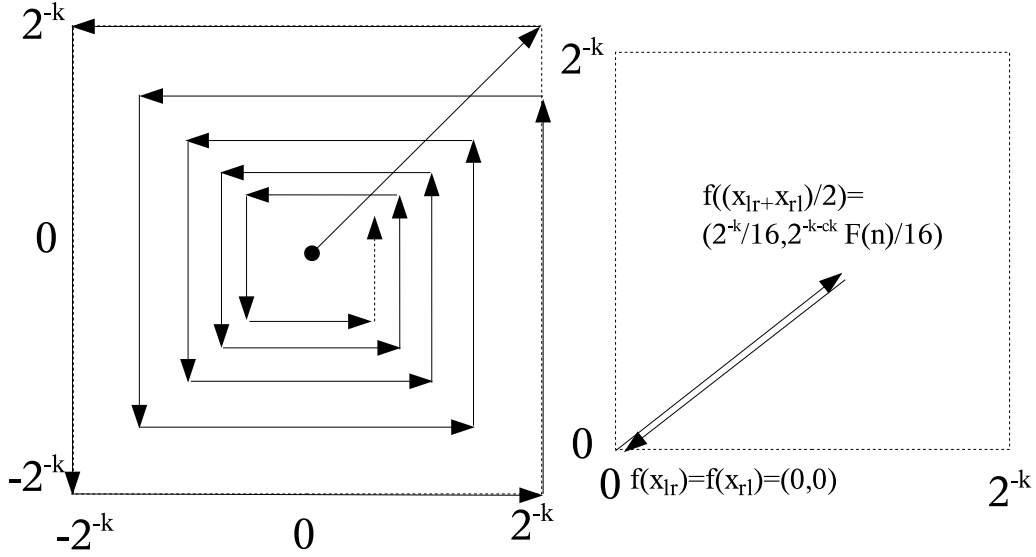

Figure 8: The construction of f

First of all we find the interval $I_n$ which is relevant to the particular $d$. If there is more than one relevant $I_n$ then the problem easily decidable in poly-time. Next, it is easy to decide whether $d$ is close to the graph of $f$ on the intervals $L, CL, CR$ and $R$. The hard part is to deal with the interval $C$. Since the picture is symmetric with respect to the middle of $C$, we assume without loss of generality that $z$ is to the left from the middle of $C$.

If $z < x_{lr} - 2 \cdot 2^{-m}$, then the pixel in question is far from $C$, and the value of $f$ on $C$ is irrelevant. Otherwise, there are two cases:

**Case 1:** $x_{lr} - 2 \cdot 2^{-m} < z < x_{lr} + 3 \cdot 2^{-m}$.

In this case, in for the value of $f$ on $C$ to become relevant we must have $|x|, |y| < 2^{ck} \cdot 4 \cdot 2^{-m} + 2^{-m}$ and $|x|, |y| < 2^{-k}$. By property (c) of $f$ on $CL$ there is a point $p = (z', f(z'))$ such that $|z' - x_{lr}| < 2^{-m}$ and $|f(z') - (x, y)| < 2^{-m}$, thus $|(z', f(z')) - d| < 5 \cdot 2^{-m}$, and it is safe to output 1 in this case regardless of the value on $C$.

**Case 2:** $z > x_{lr} + 3 \cdot 2^{-m}$.

In this case we only need to care about potential points $f(z')$ on the graph with $z' > z - 2 \cdot 2^{-m} > x_{lr} + 2^{-m}$. It is not hard to see that for these values the distance between any two potential points on the graph of $f$ (for two different possible values of $F(n)$) is at least $2^{-m}/\sqrt{2}$, hence $(x, y)$ can be only close to at most 4 potential graphs of $f$, and to confirm those we need to query $F(n) = t$? for 4 $t$'s, which is done in poly-time. This shows that $\Gamma_f$ is poly-time computable. ∎

What we need is a complexity definition that extends part (b) in theorem 17 to all the graph-computable functions.

**Definition 18** *We say that a bounded multi-valued function $f : D \subset \mathbb{R}^k \to \mathbb{R}^\ell$ for some bounded computable $D$ is **graph-computable in time** $T(n, m)$ if there is an oracle Turing machine $M^\phi(m, d, n)$ which, given an oracle $\phi$ for $x \in D$, computes a function from the family*

$$M^\phi(m, d, n) = \begin{cases} 1, & \text{if } |d - y| < 2^{-n} \text{ for some } y \in f(x) \\ 0, & \text{if } |d - y| \geq 2 \cdot 2^{-n} \text{ for all } y \in f\left(B(x, 2^{-m})\right) \\ 0 \text{ or } 1 & \text{otherwise} \end{cases} \quad (7)$$

*and the computation time is bounded by $T(m, n)$. One can think about $M^\phi(m, \bullet, n)$ as (strongly) computing a set $A_m \subset \mathbb{R}^\ell$ with precision $2^{-n}$ such that*

$$f(x) \subset A_m \subset f\left(B(x, 2^{-m})\right) \quad (8)$$

The set $A_m$ can be thought of as a "vertical cross-section" of the graph of $f$. In the case of a continuous single-valued $f$, $A_m \to \{f(x)\}$ as $m \to \infty$. More generally, one can easily see that (8) is equivalent to the condition

$$\bigcap_{i=0}^{\infty} f\left(B(x, 2^{-i})\right) \subset A_m \subset f\left(B(x, 2^{-m})\right). \quad (9)$$

For definition 18 to make sense, we need it to be equivalent to definition 15 as far as *computability* is concerned.

**Theorem 19** *A function $f : D \subset \mathbb{R}^k \to \mathbb{R}^\ell$ for some closed and bounded computable $D$ is computable as per definition 18 if and only if its graph is computable (that is, it is computable as per definition 15).*

**Proof:** $\underline{f \text{ is computable as per definition } 18 \Rightarrow \Gamma_f \text{ is computable.}}$

We show that in this case $\Gamma_f$ is weakly computable. Given a point $(x, y)$ by an oracle, we run the $M^\phi(n + 2, \bullet, n + 2)$ computing $f$ on $x$ with precision $2^{-(n+2)}$. We respond to all the oracle queries of $M^\phi$ with an answer which is consistent with $x$ as well as some point in $D$. If at any stage of the computation it is impossible, then we've discovered that $x \notin D$, and we output 0. Otherwise, either $x \in D$ and we denote $x' = x$, or there is an $x' \in D$ with $|x - x'| < 2^{-(n+2)}$, such that the run of the machine on $x$ and $x'$ is the same (we don't actually need to know $x'$). In either case, we obtain a set $A_{n+2}$ satisfying condition (8) (with $n + 2$ instead of $m$ and $n$). If $d(A_{n+2}, y) < 2^{-(n+2)}$, output 1; if $d(A_{n+2}, y) > 2^{-(n+1)}$, output 0.

If $x \in D$ and $y \in f(x)$, then $y \in A_{n+2}$ by (8), so $d(A_{n+2}, y) = 0$, and we output 1.

If we output 1, then $d(A_{n+2}, y) < 2^{-(n+1)}$ thus by (8) there is a $z \in D$ such that $|x' - z| < 2^{-(n+2)}$, and $|y - f(z)| < 2^{-(n+1)}$. Thus $|(x, y) - (z, f(z))| \leq |x - z| + |y - f(z)| \leq |x - x'| + |x' - z| + |y - f(z)| < 2^{-(n+2)} + 2^{-(n+2)} + 2^{-(n+1)} = 2^{-n}$. Hence in the case $d((x, y), \Gamma_f) > 2^{-n}$, we will output 0.

$\underline{\Gamma_f \text{ is computable} \Rightarrow f \text{ is computable as per definition } 18.}$

Note that by theorem 17 the reduction in this direction cannot be too efficient. We denote by $\pi_\ell$ the projection $\pi_\ell : \mathbb{R}^k \times \mathbb{R}^\ell \to \mathbb{R}^\ell$, $\pi_\ell(x, y) = y$.

Given a point $x \in D$ and a triple $(m, d, n)$, $d \in \mathbb{D}^\ell$ we would like to check whether $d$ is $2^{-n}$-close to an $A_m$ as in (8). To do this we take a dyadic approximation $d_x \in \mathbb{D}^k$ of $x$ such that $|d_x - x| < 2^{-(m+2)}$. If $d$ is $5 \cdot 2^{-(n+2)}$-close to $\pi_\ell \left( \Gamma_f \cap B(d_x, 2^{-(m+2)}) \times \mathbb{R}^\ell \right)$ we output 1, if it is $7 \cdot 2^{-(n+2)}$-far from $\pi_\ell \left( \Gamma_f \cap B(d_x, 2^{-(m+1)}) \times \mathbb{R}^\ell \right)$ we output 0. This can be done by "drawing" a portion of the graph using queries for $\Gamma_f$.

If $|d - f(x)| < 2^{-n}$ for one of the values of $f(x)$ (in which case we must output 1), then $f(x) \in \pi_\ell \left( \Gamma_f \cap B(d_x, 2^{-(m+2)}) \times \mathbb{R}^\ell \right)$, and $|d - f(x)| < 5 \cdot 2^{-(n+2)}$, hence we output 1.

If we output 1, then there is a $z \in B(d_x, 2^{-(m+1)})$ such that $|f(z) - d| < 7 \cdot 2^{-(n+2)}$. We have $|z - x| \leq |z - d_x| + |d_x - x| < 3 \cdot 2^{-(m+2)}$. Thus $|f(z) - d| < 2 \cdot 2^{-n}$ for some $z \in B(x, 2^{-m})$, and we are allowed to output 1 in this case. ∎

We will later see that definition 18 extends the standard complexity definition for the case of the continuous single-valued functions. We start off with some examples from the discontinuous/multi-valued functions.

## 5.4 Examples of Function Complexity

We begin with the simplest discontinuous function, the step function $s_0(x) = 0$ if $x < 0$, and $s_0(x) = 1$ if $x \geq 0$. It is now easy to see that $s_0$ is graph-computable in linear time (under definition 18). We first query for $x$ with precision $2^{-(n+1)}$. If $|x| > 2^{-(n+1)}$ we output either $\{0\}$ or $\{1\}$, depending on the sign of $x$. If $|x| < 2^{-n}$, we output $\{0, 1\}$, so in either case (8) is satisfied.

In general, if we have a piecewise continuous single-variable function $f$ with finitely many pieces $g_0, g_1, \ldots, g_k$ and computable discontinuity points $a_1, a_2, \ldots, a_k$, then $f$ is computable, and the time complexity $T(f)$ satisfies

$$T(f) \leq \sum_{i=0}^{k} T(g_i) + \sum_{i=1}^{k} T(a_i) + O(n).$$

Similar relations can be derived for functions over higher dimensions.

Consider the square root function $\sqrt{\phantom{x}} : \mathbb{C} \to \mathbb{C}$. It is two-valued at all points but 0. It cannot be made computable on $\mathbb{C}$ in the bit-sense because it does not have a continuous branch defined on the entire $\mathbb{C}$. It is also uncomputable in the original BSS model, but has an efficiently computable branch in $BSS_\mathbb{C}^{\varepsilon, b}$.

The following algorithm graph-computes the square root in poly-time on a bounded domain:

On an input $x$, $(m, d, n)$:

1. Check whether $|x| < 2^{-(2n+4)}$. In this case $f(x)$ with precision $2^{-n}$ looks like the point $\{0\}$.

2. Otherwise, we can compute $r$ and $\phi$ such that $x = re^{2\pi i \phi}$ with precision $2^{-\Omega(n+m)}$ in poly-time,

3. compute $r' = \sqrt{r}$, the positive real root,

4. take one of the values of $\phi' = \frac{\phi}{2} \pmod{1}$,

5. we take $A_m = \{r'e^{2\pi i\phi'}, -r'e^{2\pi i\phi'}\}$.

It is not hard to see that other simple multi-valued functions, such as $x \mapsto \sqrt[k]{x}$ on the complex plane are computable in poly-time.

Next, we consider the characteristic function $\chi_A$ for a bounded $A \subset \mathbb{R}^n$. Note that the computability of $\chi_A$ is equivalent to the computability of $\Gamma_{\chi_A}$, which is equivalent to the computability of $A$ and $A^c$.

Consider the complexity of $\chi_A$. On an input $x$ we need to output one of the three possible sets: $\{0\}$, $\{1\}$ or $\{0, 1\}$. We must include 1 if $x \in A$, and must exclude it if $x$ is $2^{-m}$-far from $A$. Similarly, we must include 0 if $x \notin A$, and must exclude it if $x$ is $2^{-m}$-far from $A^c$. We see that the complexity of $\chi_A$ in this case is roughly equal to the sum of the *weak* complexities of $A$ and $A^c$.

## 5.5 The Definition Extends Standard Function Complexity

In this section we show that for continuous functions with a reasonably small modulus of continuity graph-complexity extends the standard complexity definition. In particular such a function is poly-time computable if and only if it is poly-time computable according to definition 18. We prove the following theorem:

**Theorem 20** *Let $f : D \to \mathbb{R}^k$ be a continuous function, where $D \subset \mathbb{R}^\ell$ is bounded. Then the following holds:*

1. *If $f$ is computable in time $T(n)$ according to the standard definition, then it is computable in time $T(n+2) + O(n)$ according to definition 18.*

2. *If $f$ is computable in time $S(m, n)$ according to definition 18, and the modulus of continuity for $f$ is a computable $\mu = \mu(n)$ (so that $|f(x) - f(y)| < 2^{-n}$ whenever $|x - y| < 2^{-\mu(n)}$), then $f$ is computable in time $O(n \cdot S(\mu(n+2), n+2))$ according to the standard definition.*

**Proof:** (1) For an input $(m, d, n)$, compute an approximation $q$ of $f(x)$ with precision $2^{-(n+2)}$. If $|d - q| < 5 \cdot 2^{-(n+2)}$, output 1. If $|d - q| > 7 \cdot 2^{-(n+2)}$ output 0. It is not hard to see that the result of such computation satisfies (7). The time required here is $T(n+2) + O(n)$ for the comparisons.

(2) The set $A_{\mu(n+2)} \subset f(B(x, \mu(n+2)))$ is contained in $B(f(x), 2^{-(n+2)})$, and we can approximate $f(x)$ within $2^{-n}$ by $O(n)$ queries about the distance of $d$ from this set in different scales. This is done in time $O(n \cdot S(\mu(n+2), n+2))$. ∎

In particular, theorem 20 implies that a continuous function with a polynomial modulus of continuity is poly-time computable if and only if it is poly-time computable according to definition 18.

## 5.6 Different Bit-Complexity Notions for Discontinuous Functions

In this section we compare the three different complexity notions for general (possibly discontinuous and multi-valued) functions. In theorem 17 we have seen three poly-time computability notions, in the general case they correspond to the following:

(a) The graph $\Gamma_f$ is a poly-time computable as a set.
(b) $f$ is a poly-time computable function as per definition 18.
(c) The graph $\Gamma_f$ is weakly poly-time computable as a set.

In the continuous case we have seen that $(a) \Rightarrow (b) \Rightarrow (c)$, while the converses are extremely unlikely. In the general case we have:

**Theorem 21** *In the case of a general $f$:*

1. $(a) \Rightarrow (c)$ and $(b) \Rightarrow (c)$,

2. *there is a function that satisfies (b) and (c), but does not satisfy (a) unless $P = NP$,*

3. *there is a function that satisfies (a) and (c), but does not satisfy (b) unless $P = NP$ (FACTORING $\in P$ in the single-valued case).*

**Proof:** The first two parts are standard constructions, and are left to the reader. We will only prove the third part.

The proof uses the same idea as the construction in theorem 17, but is much simpler due to the fact that we are now allowed multi-valued, discontinuous functions. The multi-valued function we construct is $f : [0,1] \to P([0,1])$.

Subdivide the interval $[0,1]$ into the intervals $I_j = \left[\frac{1}{j+1}, \frac{1}{j}\right]$. For each $j \in \mathbb{N}$ set $f(\frac{1}{j}) = [0,1]$, also set $f(0) = [0,1]$.

Let $n$ be some natural number. If $n$ does not encode a boolean formula, set $f(x) = \{0\}$ for all $x \in \left[\frac{1}{n+1}, \frac{1}{n}\right]$. Otherwise, $n$ encodes a boolean formula $\phi(\vec{x})$, where $\vec{x}$ is a vector of $k < \log n - 1$ variables.

We create a $1 - 1$ correspondence between points of the form $\frac{1}{2} + \frac{t}{2^{k+1}} \in \left[\frac{1}{2}, 1\right]$, $t = 0, 1, \ldots, 2^k - 1$ and the possible assignments for $\phi$. For any $x \in \left[\frac{1}{n+1}, \frac{1}{n}\right]$, set

$$f(x) = \{0\} \cup \left\{\frac{1}{2} + \frac{t}{2^{k+1}} : t \text{ corresponds to a satisfying assignment for } \phi\right\}.$$

A graph of $f$ is schematically presented on figure 9.

We show that $\Gamma_f$ is (strongly) poly-time computable. Let $(d_1, d_2) \in \mathbb{D}^2$ and $n$ be given. We need to decide whether $(d_1, d_2)$ is $2^{-n}$-close or $2 \cdot 2^{-n}$-far from $(d_1, d_2)$. If $d_1 \notin [0,1]$, or $d_2 \notin [0,1]$ it is very easy to answer the query. Suppose $d_1 \in \left[\frac{1}{m+1}, \frac{1}{m}\right]$ for some $m$. If $\frac{1}{m} < 2^{-n}$ we can just output 1, and we do not need to compute $f$. Otherwise, we check whether $m$ corresponds to a boolean formula $\phi$ in $k$ variables. If it doesn't, then it is very easy to answer the query. Otherwise, since $2^{-n} < \frac{1}{m} < \frac{1}{2^{k+1}}$, we will need to make at most two substitutions to $\phi$ in order to correctly answer the query.

On the other hand, if $f$ were graph poly-time computable as in (b), we would be able to decide satisfiability in poly-time, which would imply that $P = NP$. ∎

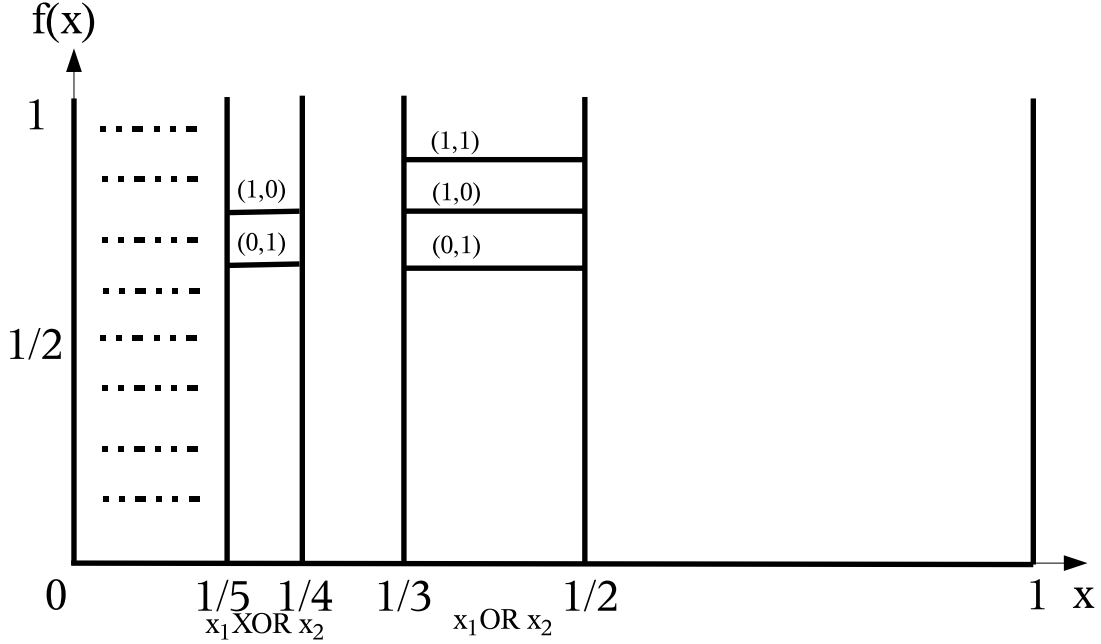

Figure 9: The construction of f

## 5.7 Notes on Complexity in the BSS Model

In theorem 16 we have seen that from the *computability* point of view, graph computability extends $BSS_C^{\varepsilon,b}$-computability of functions. In this section we give an example that illustrates the gap between the definition from the *complexity* point of view.

We give an example of a set $A \subset [0,1] \times [-1,1]$ which is $BSS$ computable by a machine $M$ without constants or error, such that the running time of $M$ on $(x,y)$, $0 < x \leq 1$ is bounded by a polynomial in $-\log x$. Thus the function

$$f(x,y) = x \cdot \chi_A(x,y)$$

is $BSS_C^{\varepsilon,b}$ computable, with a time bound polynomial in $-\log \varepsilon$ (this is the natural "poly-time computability" for this model). On the other hand, graph-computing $f$ in poly-time would allow us to solve SAT in $P$, and imply that $P = NP$.

The set $A$ is defined as follows: it is a union of $A_i$'s which are small sets in the neighborhood of $\left(\frac{1}{i}, 0\right)$, and $i = 2s$, where $s$ is a number representing a boolean formula $\phi$ with $k < \log i$ variables. We associate the $2^k$ assignments to $\phi$ with numbers from 0 to $2^k - 1$. $A_i$ is defined as follows:

$$A_i = \left\{ (x,y) \ : \ \frac{1}{i} < x < \frac{1}{i-1},\ y \geq 0,\ 0 < \frac{y}{x - 1/i} < 1,\ \text{and } \phi\left(\left\lfloor \frac{2^k \cdot y}{x - 1/i} \right\rfloor\right) = 1 \right\}.$$

On figure 10 is an illustration of $A_i$ corresponding to $\phi(x_1, x_2, x_3) = (x_1 \wedge x_2) \vee x_3$.

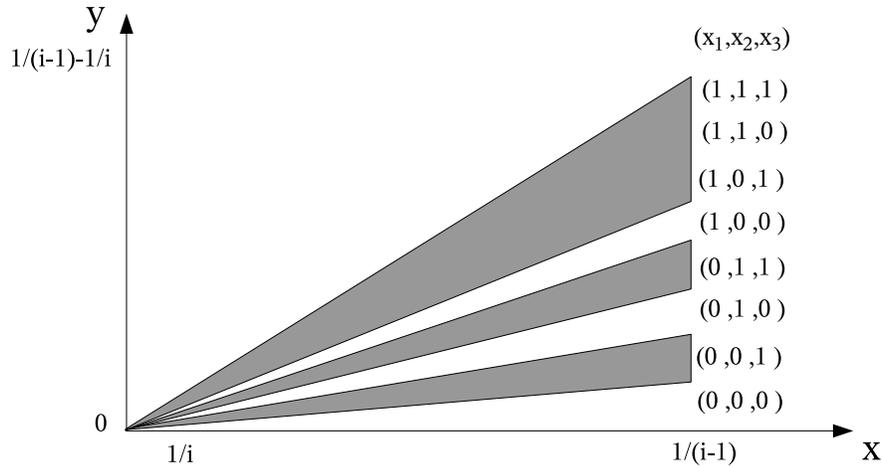

Figure 10: $A_i$ corresponding to $\phi(x_1, x_2, x_3) = (x_1 \wedge x_2) \vee x_3$

To compute $A$ in the BSS model one first locates the $i$ such that $\frac{1}{i} < x < \frac{1}{i-1}$. This takes time linear in $-\log x$. Then it takes time polynomial in $\log i \approx -\log x$ to verify whether $(x, y) \in A_i$. Thus $A$ is BSS computable in time polynomial in $-\log x$.

Now suppose that $f(x, y)$ is poly-time computable in the graph model. Then a run on $x = \frac{1}{2s}$, $y = 0$ with $2^{-m} < \frac{1}{2s^2}$ will reveal whether or not the formula $\phi$ corresponding to $s$ is satisfiable, yielding a poly-time algorithm for SAT.